\input{aipcheck}

\documentclass[
    ,final            
  ]
  {aipproc}

\layoutstyle{6x9}

\begin{document}

\title{Radiation Hydrodynamics of 
Line-Driven Winds}


\author{Stan Owocki}{
  address={Bartol Research Institute, Department of Physics and
  Astronomy \\ University of Delaware, Newark, DE 19716 USA}
}

%
%

\begin{abstract}
Dimtri Mihalas' textbooks in the 70's and 80's on {\it Stellar
Atmospheres} and {\it Foundations of Radiation Hydrodynamics} helped
lay the early groundwork for understanding the moving atmospheres and
winds of massive, luminous stars. Indeed, the central role of the
momentum of stellar radiation in driving the mass outflow makes such
massive-star winds key prototypes for radiation hydrodynamical
processes. This paper reviews the dynamics of such radiative driving,
building first upon the standard CAK model, and then discussing
subtleties associated with 
the development and saturation of instabilities, and
wind initiation near the sonic point base.
An overall goal is to illuminate the rich
physics of radiative driving and the challenges that lie ahead in
developing dynamical models that can explain the broad scalings of
mass loss rate and flow speed with stellar properties, as well as the
often complex structure and variability observed in massive-star
outflows.
%
%
%
\end{abstract}

\maketitle


\section{Introduction}

Through his textbooks and numerous  papers, Dimitri Mihalas helped
lay the groundwork for {\em radiation hydrodynamics} as an active
discipline within astrophysics.
Many of the contributions to this conference in his honor have
emphasized the common roles that radiation plays both as a diagnostic
and as an energy source/sink for astrophysical fluids.
But perhaps the most fundamental aspect of radiation hydrodynamics
arises when the {\em momentum} of radiation plays the central role in 
the dynamical driving of an astrophysical flow.
A key prototype for this lies in the strong stellar winds from hot, massive,
luminous stars, which are driven by the scattering of the star's continuum
radiation flux by line-transitions of metal ions.

The discussion here aims to summarize some basic physical concepts from 
the nearly 4 decades of research since line-driving was first proposed
as the mechanism for hot-star winds \cite{LS70}.
I begin with a summary of the standard CAK/Sobolev
\cite{CAK,So60} formalism for spherical, 
steady-state models of such winds.
I then review subsequent studies that emphasize how relaxation of the 
Sobolev approximation for localized line-transport is central 
to modeling both the strong, small-scale instability of line-driving, 
as well as the wind initiation near the transonic wind base.
A subtle but important aspect of this is the strong, dynamical role
of the {\em diffuse} component of line-scattered radiation, 
which is entirely ignored in the standard CAK/Sobolev approach,
but which can be approximately accounted for through 
{\em nonlocal, integral} forms for the escape probability.

\section{The CAK/Sobolev Model for Steady Winds}

Consider a steady-state stellar wind outflow in which 
the radiative acceleration $g_{rad}$ overcomes the 
local gravity $ GM_{\ast}/r^{2}$ at radius $r$ to drive a net acceleration
$v (dv/dr)$ in the radial flow speed $v(r)$. 
Since overcoming gravity is key, it is convenient to define a dimensionless 
equation of motion that scales all  accelerations by gravity,
\begin{equation}
(1-w_{s}/w) \, w' = -1 + \Gamma_{rad} 
\, ,
\label{dimlesseom}
\end{equation}
where $\Gamma_{rad} \equiv g_{rad} \, r^{2}/GM_{\ast}$,
$w \equiv v^{2}/v_{esc}^{2}$,
and $w' \equiv dw/dx$,
with  $x \equiv 1-R_{\ast}/r$ and $v_{esc} \equiv \sqrt{2GM_{\ast}/R_{\ast}}$
the escape speed from the stellar surface radius $R_{\ast}$. 
Eqn.\ (\ref{dimlesseom}) neglects gas pressure terms on the right side, 
since for isothermal sound speed $a$
these are of order $w_{s} \equiv (a/v_{esc})^{2} \approx 0.001$
compared to competing terms  needed to drive the wind.

For pure electron scattering opacity $\kappa_{e}$, the scaled radiative acceleration 
is just the usual Eddington parameter \cite{ASE26},
\begin{equation}
    \Gamma_{e} \equiv \frac{\kappa_{e} L_{\ast}}{4\pi GM_{\ast}c} 
     = 2 \times 10^{-5} \frac{L_{\ast}/L_{\odot}}{M_{\ast}/M_{\odot}} 
     \, .
\label{gamedddef}
\end{equation}
Because typically $L_{\ast} \sim M_{\ast}^{3-3.5}$, stars with 
$M_{\ast} > 10 M_{\odot}$ have $\Gamma_{e} > 10^{-3}$, with the
Eddington limit $\Gamma_{e} \rightarrow 1$ perhaps even being central
to setting a stellar upper mass limit of $M_{\ast} \le 150 M_{\odot}$
\cite{F05}.
Eruptive mass loss from luminous blue
variable (LBV) stars like $\eta$~Carinae might in fact be continuum-driven during
episodes of super-Eddington luminosity \cite{DH97, OGS04, OvM08}.

\begin{figure}
\includegraphics[width=14.5cm]{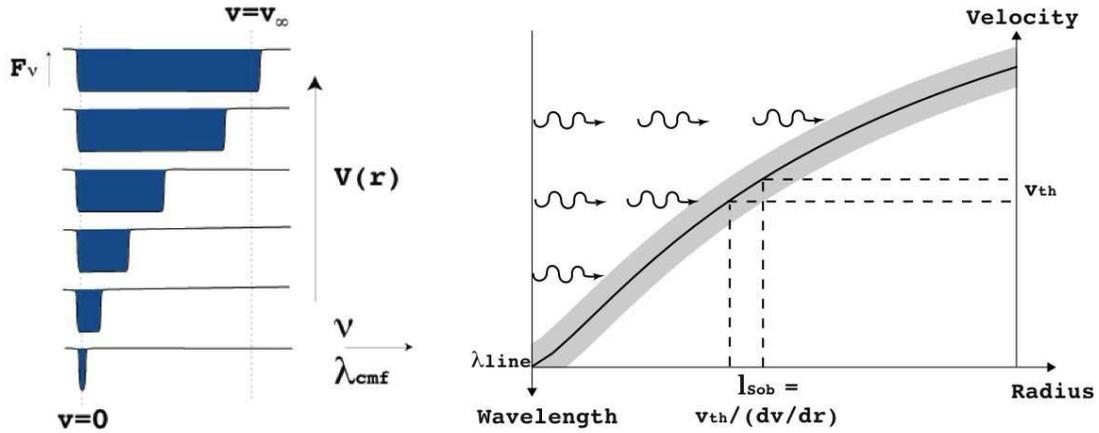}
\caption{
Two perspectives for 
the Doppler-shifted line-resonance in an accelerating flow. 
Right:
Photons with a wavelength just shortward of a line
propagate freely from the stellar surface up to a layer
where the wind outflow Doppler shifts the line into a resonance over a 
narrow width (represented here by the shading) equal to the Sobolev length,
set by the ratio of thermal speed to velocity gradient, 
$l_{Sob} \equiv v_{th}/(dv/dr)$. 
Left: 
Seen from successively larger radii within the accelerating wind, 
the Doppler-shift sweeps out an increasingly broadened line
absorption trough in the stellar spectrum.
}
\label{fig1}
\end{figure}

But the resonant nature of line (bound-bound) scattering from metal ions
leads to an opacity that is inherently much stronger than from free electrons.
For example, in the somewhat idealized, {\it optically thin} limit that all 
the line opacity could be illuminated with a flat, unattenuated 
continuum spectrum with the full stellar luminosity, the total line-force 
would exceed the free-electron force by a huge factor, of order 
${\overline Q} \approx 2000 $
\cite{Ga95}.
This implies line-driven winds can be initiated in 
even moderately massive stars with $\Gamma_{e} > 5 \times 10^{-4}$,
while for more massive stars with 
$\Gamma_{e} \approx 1/2$,
the net outward line acceleration in principle could be as 
high as $\Gamma_{thin} \approx {\overline Q} \Gamma_{e} \approx 1000$ 
times the acceleration of gravity!

In practice, self-absorption within strong lines limits the
acceleration, with the mass loss rate ${\dot M}$ set at the level 
for which the line driving is just sufficient to overcome gravity.
Indeed line-saturation keeps the dense, nearly static layers of the atmosphere
gravitationally bound.
But as illustrated by figure \ref{fig1}, within the accelerating wind, the Doppler shift 
of the line-resonance out of the absorption shadow of underlying material 
exposes the line opacity to a less attenuated flux.
This effectively desaturates the lines by limiting the resonance to a
layer with width set by the Sobolev length, $l_{Sob} = v_{th}/(dv/dr)$, 
and with optical depth proportional to
$t_{} \equiv \kappa_{e} \rho c /(dv/dr) $ 
$= \Gamma_{e} {\dot M} c^{2}/L_{\ast} w^{\prime}$.

For the CAK line-ensemble with a power-law number distribution in line-strength,
the cumulative force is reduced
by a factor  $1/({\overline Q}t)^{\alpha}$ from the optically thin 
value,
\begin{equation}
\Gamma_{CAK}
= { {\overline Q} \Gamma_{e} \over (1-\alpha) ({\overline Q} t_{} )^{\alpha} } \, 
= \Gamma_{e} k t^{-\alpha} = C (w')^{\alpha} \, ,
\label {gamcak}
\end{equation}
where the second equality defines the CAK ``force muliplier''
$kt^{-\alpha}$, with\footnote{
Here we use a slight variation of the standard CAK notation in which
the artificial dependence on a fiducial ion thermal speed is avoided by
simply setting $v_{th} = c$.
Backconversion to CAK notation is achieved by multiplying
$t$ by $v_{th}/c$ and $k$ by $\left ( v_{th}/c \right)^{\alpha}$.
The line normalization ${\overline Q}$ 
offers the advantages of being a dimensionless measure of line-opacity 
that is independent of the assumed ion thermal
speed,  with a nearly constant characteristic value of order
${\overline Q} \sim 10^3$ for a wide range of ionization conditions
\cite{Ga95}.} 
$k \equiv {\overline Q}^{1-\alpha}/(1-\alpha)$.
The last equality relates the line-force to the flow acceleration,
with
\begin{equation}
C \equiv
\frac {1}{1-\alpha} \, \left [ \frac {L_{\ast}} {{\dot M} c^2} \right ]^\alpha \,
\left [ {\overline Q} \Gamma_{e} \right ]^{1-\alpha} \, .
\label {cdef}
\end{equation}
Note
that, for fixed sets of parameters for the star ($L_{\ast}$, $M_{\ast}$, 
$\Gamma_{e}$) and
line-opacity ($\alpha$, ${\overline Q}$), this constant scales with the mass loss rate 
as $C \propto 1/{\dot M}^{\alpha}$.

Neglecting the small sound-speed term $w_{s} \approx 0.001 \ll 1 $,
application of eqn.\ (\ref{gamcak}) into (\ref{dimlesseom}) gives 
the CAK equation of motion,
\begin{equation}
F = w' + 1 - \Gamma_{e} - C (w')^\alpha = 0 \, .
\label{cakeom}
\end{equation}
For small ${\dot M}$ (large $C$), there are two solutions, while for
large ${\dot M}$ (small $C$), there are no solutions.
The CAK critical solution corresponds to a {\it maximal} mass loss rate,
defined by $\partial F/\partial w' = 0$, for which the $C(w')^{\alpha}$ is
tangent to the line $1-\Gamma_{e} + w'$ at a critical acceleration
$w'_{c} = (1-\Gamma_{e}) \alpha/(1-\alpha )$.
Since the scaled equation of motion (\ref{cakeom}) has no explicit 
spatial dependence, this critical acceleration  applies throughout 
the wind, and so can be trivially integrated to yield 
$w(x) = w_{c}'\, x$.
In terms of dimensional quantities, this represents
a specific case of the general ``beta''-velocity-law,
\begin{equation}
v(r)=
v_\infty
\left ( 1- \frac {R_{\ast}}{r} \right )^{\beta} \, ,
\label{CAK-vlaw}
\end{equation}
where here $\beta=1/2$, and
the wind terminal speed $v_\infty = v_{esc} 
\sqrt{\alpha(1-\Gamma_{e})/(1-\alpha)}$.
Similarly, the critical value 
$C_{c}$
yields, through eqn.\ (\ref{cdef}), 
the standard CAK scaling for the mass loss rate
\begin{equation}
{\dot M}_{CAK}=\frac {L_{\ast}}{c^2} \; \frac {\alpha}{1-\alpha} \;
{\left[ \frac {{\overline Q} \Gamma_{e}}{1- \Gamma_{e}} \right]}^{{(1-\alpha)}/{\alpha}} \, .
\label{mdcak}
\end{equation}

These CAK results strictly apply only under the idealized assumption that
the stellar radiation is radially streaming from a point-source.
If one takes into account the finite angular extent of the stellar disk, 
then near the stellar surface the radiative force is reduced by 
a factor $f_{d\ast} \approx 1/(1+\alpha)$,
leading to a reduced mass loss rate \cite{FA86, PPK86}
\begin{equation}
{\dot M}_{fd} = f_{d\ast}^{1/\alpha} {\dot M}_{CAK} 
= \frac{ {\dot M}_{CAK}}{(1+\alpha)^{1/\alpha}}
 \approx {\dot M}_{CAK}/2 \, .
\label{mdfd}
\end{equation}
Away from the star, the correction factor increases back toward unity,
which for the reduced base mass flux implies a stronger, more extended
acceleration, giving a somewhat higher terminal speed,
$v_{\infty} \approx 3 v_{esc}$, and a flatter velocity law,
approximated by replacing the exponent in
eqn.\ (\ref{CAK-vlaw}) by $\beta \approx 0.8$.

The effect of a radial change in ionization can be approximately taken into account
by correcting the CAK force (\ref{gamcak}) by a factor of the form
$
\left ( {n_e / W } \right )^\delta ,
$
where $n_e$ is the electron density, 
$W \equiv 0.5 ( 1-\sqrt{1-R_{\ast}/r} )$ 
is the radiation ``dilution factor'', and the exponent has a typical value 
$\delta \approx 0.1$ \cite{Ga95}.
This factor introduces an additional density dependence to that already implied
by the optical depth factor $1/t_{}^\alpha$ given in eqn.\
(\ref{gamcak}).
Its overall effect can be roughly accounted with the simple
substition $\alpha \rightarrow \alpha' \equiv \alpha - \delta$ in the power
exponents of the CAK mass loss scaling law (\ref{mdcak}).
The general tendency is to moderately increase ${\dot M}$, and accordingly to
somewhat decrease the wind speed.

The above scalings also ignore the finite gas pressure associated with a
small but non-zero sound-speed parameter $w_{s}$.
Through a perturbation expansion of the equation of motion
(\ref{dimlesseom}) in this small parameter, it possible
to derive simple scalings for the fractional corrections 
to the mass loss rate and terminal speed
\cite{OuD04},
\begin{equation}
\delta m_{s} \approx { 4\sqrt{1-\alpha} \over \alpha} \, {a \over v_{esc}}
~~~~ ;  ~~~~ \delta v_{\infty,s} 
\approx { - \alpha \delta m_{s} \over 2(1-\alpha) }
\approx {-2 \over \sqrt{1-\alpha} } \, {a \over v_{esc} } \, .
\label{dmdvs}
\end{equation}
For a typical case with $\alpha \approx 2/3$ and $w_{s}=0.001$, 
the net effect is to increase the mass loss rate and decrease the wind 
terminal speed, both by about 10\%.

Finally, an important success of these CAK scaling laws is the theoretical
rationale
they provide for an empirically observed ``Wind-Momentum-Luminosity'' 
(WML) relation \cite{Ku95}.
Combining the CAK mass-loss law (\ref{mdcak}) together with the 
scaling of the terminal speed 
with the effective escape,
we obtain a WML relation of the form,
\begin{equation}
    {\dot M} v_{\infty} \sqrt{R_{\ast}} \sim L^{1/\alpha'} 
{\overline Q}^{1/\alpha'-1} 
\end{equation}
wherein we have neglected a residual dependence on $M(1-\Gamma_{e})$ that
is generally very weak for the usual c ase that $\alpha'$ is near $2/3$. 
Note that the direct dependence ${\overline Q} \sim Z$ provides the
scaling of the WML with metalicity $Z$.

\section{Non-Sobolev Wind Instability \& Initiation}

The above CAK steady-state model depends crucially on the use of the
Sobolev approximation to compute the local CAK line force
(\ref{gamcak}).
Analyses that relax this approximation show that the flow is subject
to a strong, ``line-deshadowing instability'' (LDI) for 
velocity perturbations on a scale near and
below the Sobolev length $l_{Sob} = v_{th}/(dv/dr)$
\cite{LS70, MHR79, OR84, OR85, OP96}.
Moreover, the diffuse, scattered component of the line force, 
which in the Sobolev limit is nullified by the fore-aft symmetry of
the Sobolev escape probability (see figure \ref{fig2}b), 
turns out to have important dynamics effects on both the instability 
and sonic-point initiation of the wind.

\subsection{Linear Analysis of Line-Deshadowing Instability}

For sinusoidal perturbations ($\sim e^{i(kr-wt)}$) with wavenumber 
$k$ and frequency $\omega$, the linearized momentum equation
(ignoring the small gas pressure) relating
the perturbations in velocity and radiative acceleration implies
\begin{equation}
\omega = i \frac{\delta g}{\delta v}
\, ,
\label{6.2.2}
\end{equation}
which shows that unstable growth, with $\Im{\omega} > 0$, requires 
$\Re{(\delta g/\delta v}) > 0$.
For a purely Sobolev model \cite{Ab80}, the CAK scaling of the
line-force (\ref{gamcak}) with velocity gradient $v'$ implies
$\delta g \sim \delta v' \sim ik \delta v$, giving a purely real
$\omega$, and thus a stable wave that propagates inward at phase
speed,
\begin{equation}
\frac{\omega}{k} = - \frac{\partial g}{\partial v'} 
\equiv - U 
\, ,
\label{dgcak}
\end{equation}
which is now known as the ``Abbott speed''.
Abbott \cite{Ab80} showed this is comparable to the outward wind flow speed, and
in fact exactly equals it at the CAK critical point.

\begin{figure}
    \includegraphics[width=12cm]{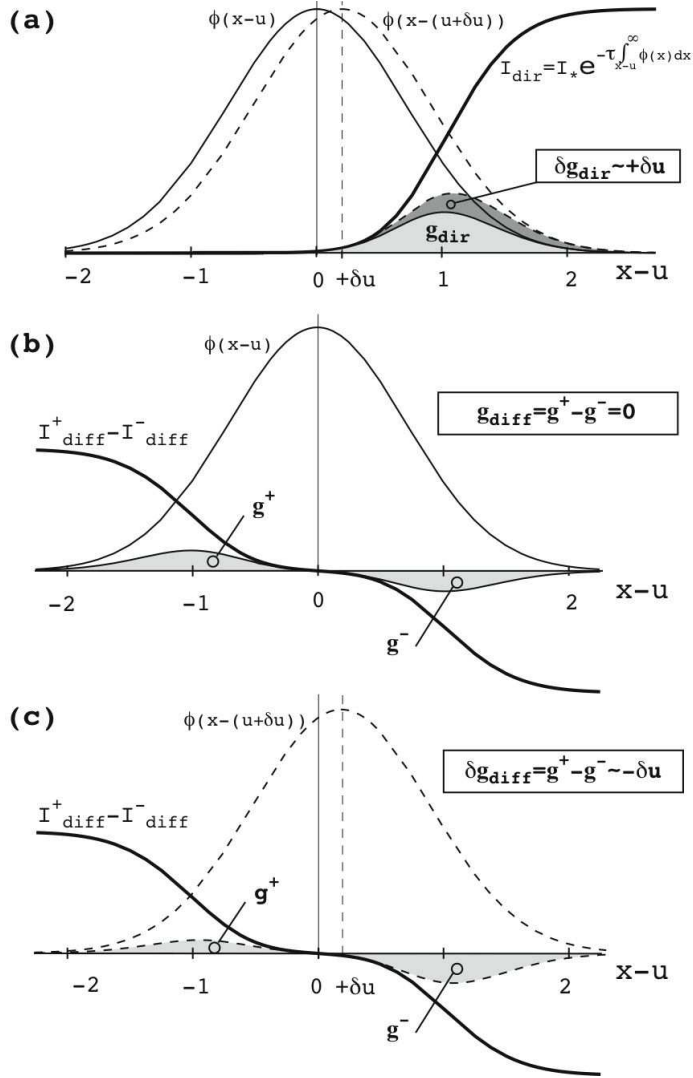}
\caption{
(a) The line profile $\phi$ and direct intensity plotted vs.
comoving frame frequency $x-u = x-v/v_{th}$, 
with the light shaded overlap area proportional 
to the net direct line-force $g_{dir}$.
The dashed profile shows the effect of the Doppler shift from a 
perturbed velocity $\delta v$, with 
the resulting extra area in the overlap with the blue-edge intensity 
giving a perturbed line-force $\delta g $ that scales in proportion to this
perturbed velocity $ \delta u = \delta v/v_{th}$.
(b) The comoving-frequency variation of the forward (+) 
and backward (-) streaming parts of
the diffuse, scattered radiation, illustrating the  cancelling of 
the overlap with the line profile that causes the net diffuse force 
to nearly vanish 
in a smooth, supersonic outflow.
(c) However, because of the Doppler shift from the perturbed velocity, 
the dashed profile now has a stronger interaction with the backward 
streaming diffuse radiation. This results in a diffuse-line-drag force 
that scales with the negative of the perturbed velocity, 
and so tends to counter the instability of the direct line-force in part a.
}
\label{fig2}
\end{figure}

As illustrated in figure \ref{fig2}a, instability arises from the deshadowing of
the line by the extra Doppler shift from the velocity perturbation,
giving $\delta g \sim \delta v$ and thus $\Im{\omega} > 0$.
A general analysis \cite{OR84} yields a ``bridging law'' encompassing both
effects,
\begin{equation}
{\delta g_{} \over \delta v} 
\approx \Omega { i k \Lambda \over  1 + ik \Lambda }
\, ,
\label{dgbridge}
\end{equation}
where 
$\Omega \approx g_{cak}/v_{th}$
sets the instability growth rate,
and the ``bridging length'' $\Lambda$ is found to be of order the 
Sobolev length $l_{sob}$.
In the long-wavelength limit $k \Lambda \ll 1$,
we recover the stable, Abbott-wave 
scalings of the Sobolev approximation,
$\delta g_{}/ \delta v  \approx i k \Omega \Lambda = ik U$;
while in the 
short-wavelength limit $k \Lambda \gg 1$,
we obtain the instability scaling
$\delta g_{}  \approx \Omega \delta v$.
The instability growth rate is very large, about the flow 
rate through the Sobolev length, $\Omega \approx v/l_{Sob}$.
Since this is a large factor $v/v_{th}$ bigger than the typical wind 
expansion rate $dv/dr \approx v/R_{\ast}$, a small perturbation 
at the wind base would, within this lineary theory,
be amplified by an enormous factor, of order 
$e^{v/v_{th}} \approx e^{100}$!

\subsection{Numerical Simulations of Instability-Generated Wind
Structure}

Numerical simulations of the nonlinear evolution require
a non-Sobolev line-force computation on a spatial grid that spans the 
full wind expansion over several $R_{\ast}$, yet resolves
the unstable structure at small scales near and below the Sobolev length.
The first tractable approach \cite{OCR88} 
focussed
on the {\em absorption} of the {\em direct} radiation from the stellar core,
accounting now for the attenuation from intervening material by
carrying out a {\em nonlocal integral} for the frequency-dependent
radial optical depth,
\begin{equation}
t(x,r) \equiv \int_{R_{\ast}}^r dr' \kappa_{e} \rho(r') 
\phi \left [ x-v(r')/v_{th} \right ]
\, ,
\label{txrad}
\end{equation}
where $\phi$ is the line-profile function, and
$x$ is the observer-frame frequency from line-center 
in units of the line thermal width.
The corresponding nonlocal form for the CAK line-ensemble force from 
this direct stellar radiation is
\begin{equation}
\Gamma_{dir} (r) = \Gamma_{e} {\overline Q}^{1-\alpha} 
\int_{-\infty}^{\infty} dx \, { \phi \left ( x-v(r)/v_{th} \right ) 
                           \over t(x,r)^{\alpha} }
\, .
\label{gamdir}
\end{equation}
In the Sobolev approximation, $t(x,r) \approx \Phi (x-v/v_{th}) t$
(where $\Phi (x) \equiv \int_{x}^{\infty} \phi(x') \, dx'$),
this recovers the CAK form (\ref{gamcak}).
But for perturbations on a spatial scale near and below the 
Sobolev length, its variation also scales in proportion to the
perturbed velocity, leading to unstable amplification.
Simulations show that because of inward nature of wave propagation
implies an anti-correlation between velocity and density variation,
the nonlinear growth leads to high-speed rarefactions that steepen 
into strong {\em reverse} shocks and compress material into dense 
clumps (or shells in these 1D models) \cite{OCR88}.

The assumption of pure-absorption was criticized by Lucy \cite{Lu84}, 
who pointed out that the interaction of a velocity perturbation with
the background, {\em diffuse} radiation from line-scattering 
results in a {\em line-drag} effect that reduces, and potentially
could even eliminate, the instability associated with the 
direct radiation from the underlying star.
The basic effect is illustrated in figure \ref{fig2}.
Panel b shows how the fore-aft ($\pm$) symmetry of the diffuse
radiation leads to cancellation of the  $g_{+}$ and $g_{-}$
force components from the forward and backward streams, as
computed from a  line-profile with frequency centered on the local
comoving mean flow.
In contrast, panel c shows that the Doppler shift associated with the
velocity perturbation $\delta v$ breaks this symmetry, and leads to
stronger forces from the component opposing the perturbation.
 
Full linear stability analyses accounting for scattering effects 
\cite{OR85, OP96}
show the fraction of the direct instability that is canceled by the
line-drag  of the perturbed diffuse force depends on the ratio of the 
scattering source function $S$ to core intensity $I_{c}$,
\begin{equation}
s = \frac{r^{2}}{R_{\ast}^{2}} \, \frac{2S}{I_{c}} 
\approx \frac{1}{1 + \mu_{\ast}}
~~~ ; ~~~   \mu_{\ast} \equiv \sqrt{1-R_{\ast}^{2}/r^{2}}
\, ,
\label{sdef}
\end{equation}
where the latter approximation applies for the optically thin form
$2S/I_{c} = 1-\mu_{\ast}$.
The net instability growth rate thus becomes
\begin{equation}
\Omega (r) \approx \frac{g_{cak}}{v_{th}}  { \mu_{\ast} (r) \over 1+
\mu_{\ast} (r) }
\, .
\label{omnet}
\end{equation}
This vanishes near the stellar surface, where $\mu_{\ast}= 0$,
but it approaches half the pure-absorption rate far from the star, 
where $\mu_{\ast} \rightarrow 1$.
This implies that the outer wind is still very unstable, with
cumulative growth of ca.\ $v_{\infty}/2v_{th} \approx 50$ e-folds.

Most efforts to account for scattering line-drag in simulations of the
nonlinear evolution of the instability have centered on a 
{\em Smooth Source Function} (SSF) approach \cite{Ow91, Fe95, OP96, OP99}.
This assumes that averaging over frequency and angle makes the
scattering source function relatively insensitive to flow
structure, implying it can be pulled out of the integral in the formal
solution for the diffuse intensity.
Within a simple {\em two-stream} treatment of the line-transport,
the net diffuse line-force then depends on the {\em difference} in the 
{\em nonlocal} escape probabilities $b_{\pm}$ associated with 
forward (+) vs. backward (-) {\em integrals} of the frequency-dependent
line-optical-depth (\ref{txrad}).
For a CAK line-ensemble, the net diffuse force can be written in a form 
quite analogous to the direct component (\ref{gamdir}),
\begin{equation}
\Gamma_{diff} (r) = \frac{\Gamma_{e} {\overline Q}^{1-\alpha}}
                          {2(1+\mu_{\ast})} \,
\left [ b_{-}(r) - b_{+}(r) \right ]
\, ,
\label{gamdiff}
\end{equation}
with
\begin{equation}
b_\pm (r) \equiv 
 \int_{-\infty}^\infty dx \,  
{ \phi(x - v(r)/v_{th} )
\over \left [ t_\pm (\pm x,r) \right ]^\alpha } \, 
\label{bpmdef}
\end{equation}
where\footnote{To account for the nonradial radiation from a 
finite-angle star, in practice SSF models carry out the forward/back
optical depth integrals along a ray with an impact parameter that 
is $R_{\star}/\sqrt{2}$ off the disk center.
A similar application for the direct force (\ref{gamdir}) gives within
a few percent the proper finite-disk correction to the CAK/Sobolev
point-star line-force (\ref{gamcak}) in a smooth flow.}
for $t_{-}$ the integral bounds in (\ref{txrad}) are now from 
$r$ to the outer radius $R_{max}$ \cite{OP96, Ow04}, and
the overall normalization for $\Gamma_{diff}$ 
assumes the optically thin source function from eqn.\ (\ref{sdef}).
In the Sobolev approximation, both the forward and backward integrals 
give the same form, viz.\ 
$t_{\pm}(\pm x,r) \approx \Phi [\pm(x-v/v_{th})] t$,
leading to the net cancellation of the Sobolev diffuse force
(figure \ref{fig2}b).
But for perturbations on a spatial scale near and below the 
Sobolev length, the perturbed velocity breaks the forward/back
symmetry (figure \ref{fig2}c), leading to perturbed diffuse force that
now scales in proportion to the {\em negative} of the perturbed velocity,
and thus giving the diffuse line-drag that reduces the
net instability by the factors given in (\ref{sdef}) and (\ref{omnet}).


\begin{figure}
\includegraphics[width=15cm]{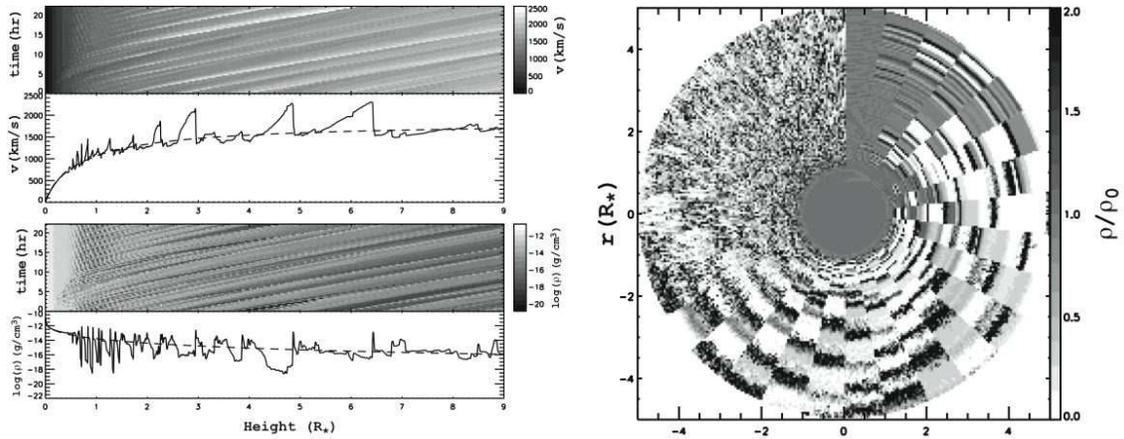}
\caption{
Left: Results of 1D Smooth-Source-Function (SSF) simulation of the 
line-deshadowing instability.
The line plots show the spatial variation of 
velocity (upper) and density (lower) at a fixed, arbitrary
time snapshot.
The corresponding grey scales show both the time (vertical axis)
and height (horizontal axis) evolution.
The dashed curve shows the corresponding smooth, steady CAK model.
Right: For 2DH+1DR SSF simulation, grayscale representation for the
density variations 
rendered as a time  sequence of 2-D wedges of the simulation model azimuthal 
range  $\Delta \phi = 12^{\rm o}$ stacked 
clockwise from the vertical in intervals of 4000~sec from the CAK
initial condition.
}
\label{fig3}
\end{figure}

The left panel of figure \ref{fig3} illustrates the results of a 1D SSF 
simulation, starting from an initial condition set by smooth, steady-state 
CAK/Sobolev model (dashed curves).
Because of the line-drag stabilization of the driving near the star 
(eqn. \ref{omnet}), the wind base remains smooth and steady.
But away from the stellar surface, the net strong instability leads to extensive 
structure in both velocity and density, roughly straddling the CAK steady-state.
Because of the backstreaming component of the diffuse line-force
causes any outer wind structure to induce small-amplitude fluctuations
near the wind base, the wind structure, once initiated, is  ``self-excited'', 
arising spontaneously without any explict perturbation from the
stellar boundary.

In the outer wind, the velocity variations become highly nonlinear and 
nonmonotonic, with amplitudes approaching $1000$~km/s, leading to 
formation of strong shocks. 
However, these high-velocity 
regions have very low density, and thus represent only very little material.
As noted for the pure-absorption models,
this anti-correlation between velocity and density arises because the 
unstable linear waves that lead to the structure have an {\it inward} 
propagation relative to the mean flow.
For most of the wind mass, the dominant overall effect of the instability 
is to concentrate material into dense clumps, which can lead to
substantial (factor 3-5) overestimates in the mass loss rate from
diagnostics (e.g.\ radio or Balmer emission) that scale with the square of
the density.


\begin{figure}
    \includegraphics[width=15cm]{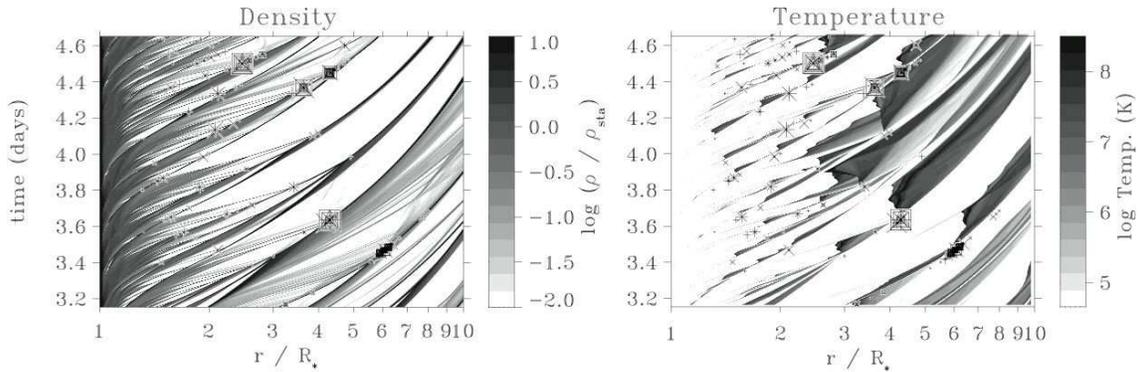}
\caption{
Greyscale rendition of the evolution of wind density and temperature,
for time-dependent wind-instability models with structure formation triggered 
by photospheric perturbations. 
The boxed crosses identify localized region of clump-clump collision that 
lead to the hot, dense gas needed for a substantial level of soft 
X-rays emission.
}
\label{fig4}
\end{figure}

The presence of multiple, embedded strong shocks suggests a potential 
source for the soft X-ray emission observed from massive star winds;
but the rarefied nature of the high-speed gas implies that this
self-excited structure actually
feeds very little material through 
the strong shocks needed to heat gas to X-ray emitting temperatures.
To increase the level of X-ray emission, 
Feldmeier \cite{FPP97}
introduced intrinsic perturbations at the wind base, assuming 
the underlying stellar photosphere has a turbulent 
spectrum of compressible sound waves characterized by abrupt phase 
shifts in velocity and density.
These abrupt shifts seed wind variations that, when amplified by the 
line-deshadowing instabilty, now include substantial velocity 
variations among the dense clumps.
As illustrated in figure \ref{fig4}, when these dense clumps 
collide, they induce regions of relatively dense, hot gas which 
produce localized bursts of X-ray emission.
Averaged over time, these localized regions can 
collectively yield X-ray emission with a brightness and spectrum that 
is comparable to what is typically observed from such hot stars.
%

Because of the computational expense of carrying out nonlocal optical depth 
integrations at each time step, such SSF instability simulations have 
generally been limited to just 1D.
More realistically, various kinds of thin-shell instabilites 
\cite{Vi94} can be expected to break up the structure 
into a complex, multidimensional form.
A first step to modelling both radial and lateral structure
\cite{DO03}
is to use a restricted  ``2D-H+1D-R'' approach, extending the 
hydrodynamical model to 2D in radius and azimuth, 
but still keeping the  1D-SSF radial integration for the inward/outward 
optical depth within each azimuthal zone.
The right panel of figure \ref{fig3} shows
the resulting 2D density structure within a
narrow ($12^{\rm o}$) wedge, with the time evolution rendered 
clockwise at fixed time intervals of 4000 sec starting from the CAK
initial condition at the top.
The line-deshadowing instability is first manifest 
as  strong radial velocity variations and associated density compressions 
that initially extend nearly coherently across the full azimuthal range of 
the computational wedge.

But as these initial ``shell'' structures are accelerated outward, 
they become progressively  disrupted by Rayleigh-Taylor or thin-shell 
instabilities  that operate in azimuth down to the grid scale 
$d \phi = 0.2^{\rm o}$.
%
Such a 2DR+1DH approach may well exaggerate the level of variation on small 
lateral scales.
The lack of {\it lateral} integration needed to compute 
an azimuthal component of the diffuse line-force means that the model 
ignores a potentially strong net lateral line-drag that should 
strongly damp azimuthal velocity perturbations on scales below the lateral
Sobolev length $l_{0} \equiv r v_{th}/v_{r}$ 
\cite{ROC90}.
Presuming that this would inhibit development of lateral instability at 
such scales, then any lateral breakup would be limited to a minimum 
lateral angular scale of $\Delta \phi_{min} \approx l_{0}/r = 
v_{th}/v_{r} \approx 0.01 \, {\rm rad} \approx 0.5^{\rm o}$.
Further work is needed to address this issue through explicit 
incorporation of the lateral line-force and the associated line-drag 
effect.

Overall, both the 1D and 2D SSF simulations suggest that the LDI results
naturally in strong, small-scale clumping, implying a significant (factor
several) reduction in mass loss rates deduced from density-squared
diagnostics. 
On the other hand, the small scale makes it unlikely individual clumps
can become optically thick, and so lead to a signifcant wind ``porosity''
that would weaken {\em single}-density processes like bound-free absorption of
X-rays \cite{OC06, OHF07}.
Together these suggest the observed weak asymmetry of X-ray emission lines
from massive stars may be mainly the result of a reduced mass loss rate, 
and not wind porosity.

\subsection{Escape Integral Source Function Method and Wind
Initiation}

Finally, a full linear analysis of the effects of scattering on the
line-deshadowing instability \cite{OR85} shows that perturbations in the line
source function can lead to unstable wavemodes that now have {\em outward}
phase propagation.
Since such outward wavemodes have a positive correlation between
density and velocity, this raises the possibility that the nonlinear
amplification of velocity perturbations might now lead to strong 
{\em forward} shocks; this could cause a much larger fraction 
of wind material to be heated to X-ray emitting temperatures, 
as needed to explain the observed X-ray emission
\cite{Pu94}.

These possibilities motivated development \cite{OP96} 
of a new {\em Escape Integral Source Function} (EISF) that now re-uses 
the nonlocal escape probabilities $b_{\pm}$ defined in 
eqn.\ (\ref{bpmdef}) to estimate the dynamical
variation of the source function,
\begin{equation}
    S(r) \approx
    I_{c} \, \left [ 1 -\mu_{\ast} (r) \right ] ~
    \frac{b_{+}(r)}{b_{+}(r) + b_{-} (r)} \, .
\label{seisfdef}
\end{equation}
Using this $S(r)$ within a formal solution for the diffuse intensity
thus provides the basis for a computation of the diffuse
line-force that accounts for both the line-drag and phase reversal
effects \cite{OP96}.

Application in an EISF simulation code \cite{OP99} indeed confirms that 
the initial linear onset of the wind instability is
dominated by variations with a {\em positive} correlation between 
density and velocity.
However, the self-shadowing of radiative driving from such denser,
faster structures makes their growth {\em saturate} at a
low amplitude, about the thermal speed $v_{th}$.
Eventually, the wind structure thus again becomes dominated by the slower, but
less-saturated growth of variations with the usual anti-correlation
between velocity and density.
Overall, the fully developed wind structure in EISF models thus
turns out to be quite similar to the SSF case, 
with no apparent tendency to produce denser, forward shocks that
might give stronger X-ray emission.

\begin{figure}
\includegraphics[width=15cm]{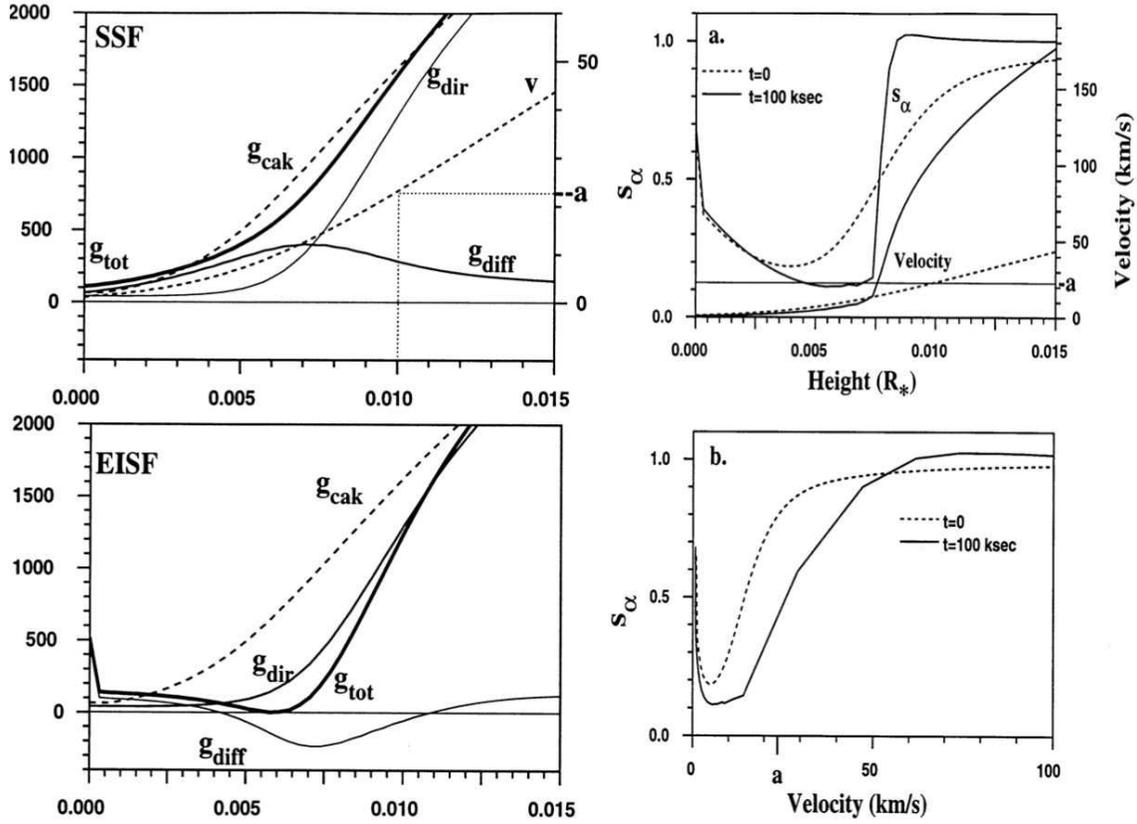}
\caption{
Left: Spatial variation of the velocity and various components of 
the line force near the transonic wind base for the SSF (top) and 
EISF (bottom) models,  computed at the CAK initial condition.
The location of the sonic point is indicated by dotted lines in the uppermost
panel.
%
Right: a. Radial variation near the transonic wind base of 
the velocity $v$  and the line-ensemble-averaged source function
$s_{\alpha}$ (normalized to the optically thin value).
The dashed curve applies at the CAK initial condition, while the solid
curve gives results for at time 100~ksec later in an EISF simulation.
b. Corresponding variations of this scaled source function $s_\alpha$ vs. 
wind velocity.
}
\label{fig5}
\end{figure}

But perhaps the biggest surprise from these EISF simulations regards the 
subtle nature of the wind initiation near the sonic point base.
Because the ion thermal speed $v_{th}$ is only moderately smaller than
the sound speed $a$, the Sobolev length $l_{Sob} = v_{th}/(dv/dr)$
near the sonic point becomes comparable to the characterize density/velocity
scale length $H \approx v/(dv/dr)$, thus raising questions about the
suitability of using the Sobolev-based CAK line-force to model the 
wind driving in this transonic region.
The left panel of figure \ref{fig5} plots the spatial variation of
various line-force components within the transonic wind based of a CAK 
initial condition, for both SSF (upper) and EISF (lower) models with a 
typical thermal to sound speed ratio $v_{th}/a = 0.28$.
Note first that in both models, the direct force lies well below the 
CAK/Sobolev force throughout the subsonic region;
as first noted by Poe et al.\ \cite{POC90}, this gives non-Sobolev
pure-absorption models a quite different character than the usual
CAK/Sobolev solution.

However, in the SSF model with scattering, there now develops 
in this region a {\em positive} diffuse force, resulting from the strong
steepening of the velocity gradient near the sonic point, which
makes $b_{-} > b_{+}$ and so through eqn.\ (\ref{gamdiff}) gives a
positive diffuse force.
Quite remarkably when added to the direct term, this gives a total force 
that follows very closely the CAK/Sobolev force throughout the subsonic
and transonic region.
This explains why the (drag-effect stabilized) base outflow in SSF
simulations follows very closely the CAK solution, and why
co-moving-frame models \cite{PPK86} of line-driven winds  
also match well the (finite-disk-corrected) CAK model. 
In effect, the addition of the SSF diffuse force corrects for the
non-Sobolev error in the pure-absorption force, making the total force
accurate to a higher order in the parameter $l_{Sob}/H$.
This suggests that the validity of a CAK/Sobolev approach, for which the 
diffuse force is normally assumed to vanish and thus play no role, actually
depends crucially on the scattering nature of line-transport for
proper wind initiation near the transonic base!

But the situation is further complicated by the EISF results, which
instead give a {\em negative} diffuse force in the subsonic region.
This stems from the strong ``dip'' in the EISF source function near
and below the sonic point, as shown in the right panels of figure
\ref{fig5}.
Much like the SSF diffuse force, 
this dip results from the velocity gradient steepening and associated 
asymmetries in the escape probabilities near the sonic point.
The lower subsonic gradient implies a lower $b_{+}$ to shift core
radiation into line resonance, while the large outward gradient leads 
to a higher $b_{-}$ that enhances the escape from the resonance zone. 
This lowers the mean intensity, and by eqn.\ (\ref{seisfdef}),
the source function.
But then the strong outward increase in mean intensity from this dip 
implies a strong {\em inward diffusion} of line-flux, giving rise to the 
inward diffuse line-force.
The net result is to weaken the total force even below its
pure-absorption value, yielding then a reduction in the mass flux that
can be driven through the sonic point.

While the EISF method does not represent a fully consistent radiative 
transfer solution, such source function dips have been seen in model
atmospheres with full NLTE solutions of the line transfer
\cite{Se93}.
It is unclear why the effect was not seen in the co-moving-frame
calculations of \cite{PPK86}, but one possibility is that this
analysis was for a relatively dense supergiant wind, for which
coupling to the continuum and thermalization effects might help keep
the line source function elevated, and smooth, in the sonic region.
If true, this would imply that the SSF method and results might be
quite appropriate for dense winds from OB supergiants.
However, for late O and cooler main sequence stay, the effects of a
source-function dip could be significant, perhaps even playing a role 
in their ``weak wind'' problem \cite{PSNH09}.
Such effects should also be considered in the context of recent 
renewed interest in the role of atmospheric
microturbulence on the wind mass loss rate \cite{Lu07}.

In conclusion, it seems that some 40 years after the initial work on
line-driven stellar winds, both wind instability and wind initiation
remain key issues for understanding their radiation hydrodynamics.
To end with a personal note, I first learned about
stellar winds as a graduate student in Dimitri Mihalas' Spring 1977
course on ``Stellar Atmospheres'' at the University of Colorado, and
indeed even wrote my term paper on the subject.
In grading that paper, Dimitri wrote on the cover page, ``Interesting 
paper; I hope you will follow it up with some future research''.
In perusing this volume of papers from the conference in honor of his 
70th birthday, perhaps he will be amused to see that at least one ex-student
has endeavored to take his advice.

\begin{theacknowledgments}
This work was carried out with partial support by NASA
grant LTSA/NNG05GC36G, and by NSF grant AST-0507581.
\end{theacknowledgments}

\end{document}